\documentclass[twocolumn]{aastex63}
\usepackage{graphicx}
\usepackage{amsmath}
\usepackage{amssymb}
\usepackage{hyperref}

\def\gtrsim{\mathrel{\hbox{\rlap{\hbox{\lower4pt\hbox{$\sim$}}}\hbox{$>$}}}}
\def\lesssim{\mathrel{\hbox{\rlap{\hbox{\lower4pt\hbox{$\sim$}}}\hbox{$<$}}}}
\def\gtrsim{\mathrel{\hbox{\rlap{\hbox{\lower4pt\hbox{$\sim$}}}\hbox{$>$}}}}
\def\farcs{\hbox{$.\!\!^{\prime\prime}$}}
\def\farcm{\hbox{$.\!\!^{\prime}$}}

\newcommand{\mjb}{\,mJy\,beam$^{-1}$}

\begin{document}


\def\chan{{\sl CXO\ }}

\title{``The Goose'' Pulsar Wind Nebula of PSR J1016--5857: The Birth of a Plerion}

\author{Noel Klingler}
\affil{Astrophysics Science Division, NASA Goddard Space Flight Center, 8800 Greenbelt Road, Greenbelt, MD, 20771, USA}
\affil{Center for Space Sciences and Technology, University of Maryland Baltimore County, 1000 Hilltop Circle, Baltimore, MD, 21250, USA}
\affil{Center for Research and Exploration in Space Science \& Technology II (CRESST II)}
\author{Oleg Kargaltsev}
\affil{Department of Physics, The George Washington University, 725 21st Street NW, Washington, DC, 20052, USA}
\author{George G.\ Pavlov}
\affil{Department of Astronomy \& Astrophysics, The Pennsylvania State University, 525 Davey Laboratory, University Park, PA, 16802, USA}
\author{C.-Y. Ng}
\affil{Department of Physics, The University of Hong Kong, Pokfulam Road, Hong Kong}
\author{Zhengyangguang Gong}
\affil{Department of Physics, The University of Hong Kong, Pokfulam Road, Hong Kong}
\affil{Universit\"ats-Sternwarte, Fakult\"at f\"ur Physik der Ludwig-Maximilians, Universit\"at M\"unchen, Scheinerstr.\ 1, 81679 M\"unchen, Germany}
\affil{Max Planck Institute for Extraterrestrial Physics, Giessenbachstr.\ 1, 85748 Garching, Germany}
\author{Jeremy Hare}
\affil{Astrophysics Science Division, NASA Goddard Space Flight Center, 8800 Greenbelt Road, Greenbelt, MD, 20771, USA}
\affil{NASA Postdoctoral Program Fellow}

\begin{abstract}
We report the results of X-ray ({\sl CXO}) and radio (ATCA) observations of the pulsar wind nebula (PWN) powered by the young pulsar PSR J1016--5857, which we dub ``the Goose'' PWN.
In both bands the images reveal a tail-like PWN morphology which can be attributed to pulsar's motion. 
By comparing archival and new {\sl CXO} observations, we measure the pulsar's proper motion  $\mu=28.8\pm7.3$ mas yr$^{-1}$, yielding a projected pulsar velocity $v \approx440\pm110$ km s$^{-1}$ (at $d=3.2$ kpc); its direction is consistent with the PWN shape. 
Radio emission from the PWN is polarized, with the magnetic field oriented along the pulsar tail. 
The radio tail connects to a larger radio structure (not seen in X-rays) which we interpret as a relic PWN (also known as a plerion).
The spectral analysis of the {\sl CXO} data shows that the PWN spectrum softens from $\Gamma=1.7$ to $\Gamma\approx2.3-2.5$ with increasing distance from the pulsar.  
The softening can be attributed to the rapid synchrotron burn-off, which would explain the lack of X-ray emission from the older relic PWN. 
In addition to non-thermal PWN emission, we detected thermal emission from a hot plasma which we attribute to the host SNR.
The radio PWN morphology and the proper motion of the pulsar suggest that the reverse shock passed through the pulsar's vicinity and pushed the PWN to one side. 
\end{abstract}


\section{INTRODUCTION}

As a pulsar spins down, most of its rotational energy is imparted into a magnetized ultra-relativistic particle wind, whose synchrotron emission can be seen from radio to X-rays as a pulsar wind nebula (PWN; see \citealt{Reynolds2017,Kargaltsev2017a} for recent reviews). 
While X-rays come from recently-produced wind (in which the particles have not had time to cool substantially), radio emission can also reflect the distribution of particles produced earlier in the pulsar's lifetime. 
These older ``relic'' particles are more numerous than the younger X-ray-emitting ones, and may also be energetic enough to produce TeV $\gamma$-rays via Inverse Compton (IC) up-scattering of the ambient photons \citep{deJager2009,Kargaltsev2013,HESS2018a}.

For pulsars which still reside inside their progenitor supernova remnants (SNRs), if the interaction with the reverse SNR shock has already occurred, the relic PWN (also known as a plerion) may be pushed aside resulting in a TeV and/or radio source being offset from the current pulsar position \citep{Blondin2001}.    
Another possible reason for offsets between the pulsar and the older population of pulsar wind particles could be the fast  motion of the pulsar.
For pulsars outside their progenitor SNRs, the ram pressure exerted by the ISM confines the wind of the supersonically-moving pulsar into a ``tail'' behind the moving pulsar (see \citealt{Kargaltsev2017b} for a recent review).  

The sample of supersonically-moving pulsars with tails seen in both radio and X-rays is small (only J1509--5850, J1357--6429, the Mouse, the Lighthouse, and B1929+10)\footnote{See \citealt{Klingler2016a,Kirichenko2016,Klingler2018,Pavan2014,Misanovic2008}, respectively.}. 
This motivated us to perform a deeper {\sl Chandra X-ray Observatory} {\sl (CXO)} observation of PSR J1016--5857 (J1016 hereafter), since the initial short {\sl CXO} observation indicated that the pulsar is likely supersonic, with a tail seen both in X-rays and radio. 

\begin{figure*}
\includegraphics[width=1.0\hsize,angle=0]{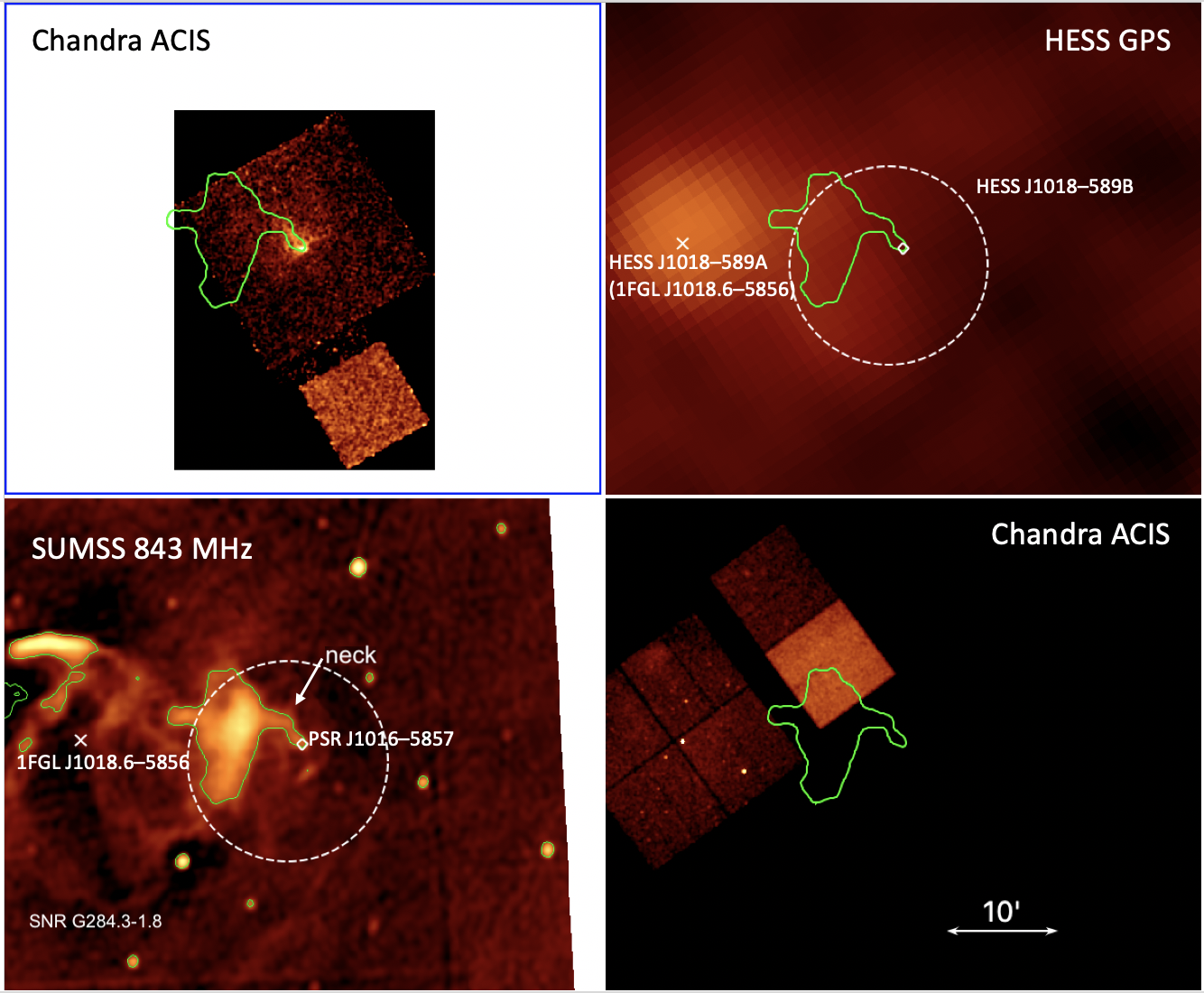}
\caption{Multiwavelength images of the J1016 field.  {\sl Top Left:} {\sl CXO} (0.5--8 keV).  {\sl Top Right:} HESS Galactic Plane Survey (0.2--100 TeV; \citealt{HESS2018b}).  {\sl Bottom Left:} SUMSS Galactic Plane Survey (843 MHz; \citealt{Green2014}).  {\sl Bottom Right:} {\sl CXO} observation of 1FGL J1018.6--5856 (0.5--8 keV).  The green contours mark the bright radio structure (possibly associated with the J1016 SNR), and the dashed white circle marks HESS J1018--589B.}
\label{fig-multiwavelength}
\end{figure*}

J1016 was discovered by the Parkes telescope in the Pulsar Multibeam Survey \citep{Manchester2001}, and was subsequently found to coincide with an {\sl Einstein Observatory} X-ray source and an unidentified EGRET source, 3EG J1013--5915 \citep{Camilo2001}. 
J1016 is a young and energetic pulsar, with rotation period $P=107$ ms, characteristic age $\tau=P/2\dot{P}=21$ kyr, and spin-down energy loss rate $\dot{E}=2.6\times10^{36}$ erg s$^{-1}$, located $\approx$20$'$ west of the center of SNR G284.3--1.8 (see Figure \ref{fig-multiwavelength}, bottom-left panel).  
Its radio pulse profile is unusual, showing a single strong asymmetric peak with a bump on one side. 
J1016 was also detected by the {\sl Fermi}-LAT, which detected a $\gamma$-ray pulse profile showing an asymmetric double peak profile \citep{Abdo2013}. 
It was also observed with the {\sl Rossi X-ray Timing Explorer}, but no X-ray pulsations were found.

Radio survey images obtained with the Molonglo Observatory Synthesis Telescope (MOST) \citep{Milne1989} show a bright radio structure whose peculiar shape somewhat resembles that of a goose in flight (highlighted by the green contours in Figure \ref{fig-multiwavelength}, hence the name ``Goose PWN''), with the pulsar located at the goose's ``head'', and with a noticeable bend in the ``neck''.
Although, in projection, J1016 appears close to SNR G284.3--1.8 (see Figure \ref{fig-multiwavelength}), the recent discovery of a high-mass $\gamma$-ray binary 1FGL J1018.6--5856 \citep{Fermi2012} within the SNR called into question the  J1016/G284.3 association \citep{Williams2015}. 
However, \citet{Marcote2018} claimed that 1FGL J1018.6--5856 and SNR G284.3--1.8 can not be related due to considerations of the binary's proper motion.
Additionally, the \citet{HESS2018a} reported that the TeV source HESS J1018--589B (see Figure \ref{fig-multiwavelength}) meets all criteria for being the TeV PWN counterpart to PSR J1016 (i.e., the pulsar parameters are consistent with the offset, size, luminosity, and surface brightness of the TeV emission).

A short {\sl CXO} observation (ObsID 3855, 18.7 ks; PI F.\ Camilo) performed in 2003 revealed an X-ray PWN whose spectrum fits an absorbed power-law (PL) model with $\Gamma=1.32\pm0.25$, $N_{\rm H}=(5.0\pm1.7)\times10^{21}$ cm$^{-2}$, and a 0.8--7 keV luminosity $L_{\rm PWN}=3.2\times10^{32}$ erg s$^{-1}$ \citep{Camilo2004}. 
The dispersion measure DM $=394$ pc cm$^{-3}$ places J1016 at a distance $d=3.2$ kpc (using the Galactic electron density model of \citealt{YMW2017}), which is consistent with the observed $N_{\rm H}$.

In this paper we report the results of new {\sl CXO} observations of PSR
J1016--5857 and its PWN analyzed jointly with the archival {\sl CXO} data as well as Australia Telescope Compact Array (ATCA) radio observations. 
In Section 2 we describe the observations and data reduction. 
In Section 3 we present the results of X-ray and radio data analysis.  
The implications of our analysis are discussed in Section 4 and summarized in Section 5.

\begin{deluxetable}{lc}
\tablecolumns{9}
\tablecaption{Observed and Derived Pulsar Parameters \label{tbl-parameters}}
\tablewidth{0pt}
\tablehead{\colhead{Parameter} & \colhead{Value} }
\startdata
R.A. (J2000.0) & 10 16 21.16(1)  \\
Decl. (J2000.0) & --58 57 12.1(1)  \\
Epoch of position (MJD) & 52717  \\
Galactic longitude (deg) & 284.079  \\
Galactic latitude (deg) & --1.880  \\
Spin period, $P$ (ms) & 107.39  \\
Period derivative, $\dot{P}$ (10$^{-14}$) & 8.0834 \\
Dispersion measure, DM (pc cm$^{-3}$) & 394.5 \\
Distance, $d$ (kpc) & 3.2, 8.0  \\
Surface magnetic field, $B_s$ (10$^{12}$ G) & 3.0  \\
Spin-down power, $\dot{E}$ (10$^{36}$ erg s$^{-1}$) & 2.6  \\
Spin-down age, $\tau_{\rm sd} = P/(2\dot{P})$ (kyr) & 21 
\enddata
\tablenotetext{}{Parameters are from the ATNF Pulsar Catalog \citep{Manchester2005}.  The DM distance estimates listed correspond to those obtained using the Galactic free electron density models of \citet{YMW2017} and \citet{Cordes2002}, respectively.}
\end{deluxetable}

\section{OBSERVATIONS AND DATA REDUCTION}

\subsection{X-rays ({\sl CXO})}

We utilized both {\sl CXO} observations of J1016: the archival ObsID 3855 (18.72 ks, ACIS-S, 2003-05-25; PI: Camilo) and the new ObsID 21357 (92.86 ks, ACIS-I, 2019-09-24; PI: Klingler).
Both were taken with the Advanced CCD Imaging Spectrometer (ACIS) instrument operating in Very Faint timed exposure mode (3.24 s time resolution).

For data processing we used the {\sl Chandra} Interactive Analysis of Observations (CIAO) software package version 4.12 \citep{Fruscione2006} and the {\sl Chandra} Calibration Database (CALDB) version 4.9.2.1.
We ran {\tt chandra\_repro} on the data sets, which applies all the necessary data processing tools and applies the latest calibrations.

We produced exposure maps for both observations and created a merged exposure-map-corrected image with {\tt merge\_obs} (using the default effective energy of 2.3 keV).
All spectra were extracted using {\tt specextract} and fitted using the HEASoft package {\tt XSPEC} (v12.11.1; \citealt{Arnaud1996}).
We used the {\tt tbabs} absorption model, which uses absorption cross sections from \citet{Wilms2000}.
All images and spectra were restricted to the 0.5--8 keV range, and uncertainties listed below are at the 1$\sigma$ confidence level.
In all images, North is up and East is left.

\subsection{Radio (ATCA)}

We analyzed archival ATCA observations of the field of J1016 taken in 3, 6, 13, and 20\,cm bands.
The observation parameters are listed in Table~\ref{table:atcaobs}. 
We performed all data reduction using the MIRIAD package \citep{Sault1995}.
After flagging bad data points and standard calibration,
we discarded all 6\,km baselines to obtain a uniform \textit{u}-\textit{v} coverage and
formed radio maps using a weighting scheme developed by \citet{Briggs1995}.
We chose \texttt{robust}=$-2$ at 20\,cm, which is equivalent to uniform weighting, to suppress sidelobes.
At higher frequencies, we used slightly larger \texttt{robust} values (0--0.5) to boost the sensitivity.
These values are listed in Table~\ref{table:atcaimg}.
We deconvolved Stokes I, Q, and U images simultaneously using a maximum entropy algorithm.
The beam sizes and RMS noise of the final images at each band are listed in Table~\ref{table:atcaimg}.

\begin{deluxetable*}{lcccccc}
\tablewidth{0pt}
\tablecaption{ATCA Radio Observation Details \label{table:atcaobs}}
\tablehead{\colhead{Obs.\ Date} & \colhead{Array}&
\colhead{Wavelength}& \colhead{Center Freq.} &
\colhead{No.\ of} & \colhead{Usable Band-} & \colhead{On-source} \\
& \colhead{Config.} & \colhead{(cm)} & \colhead{(MHz)} &
\colhead{Channels\tablenotemark{a}} &
\colhead{\protect{width\tablenotemark{a}} (MHz)}
& \colhead{Time (hr)} }
\decimals
\startdata
2001 Oct 17 & EW352 & 20, 13 & 1384, 2240 & 13 & 104 & 11 \\
2001 Oct 28 & 1.5D & 20, 13 & 1384, 2496 & 13 & 104 & 12 \\
2008 Dec 29 & 750B & 6, 3 & 4800, 8640 & 13 & 104 & 13 \\
2009 Feb 12 & EW352 & 6, 3 & 4800, 8640 & 13 & 104 & 12
\enddata
\tablenotetext{a}{per center frequency.}
\end{deluxetable*}

\begin{deluxetable}{lcccc}
\tablewidth{0pt}
\tablecaption{Parameters for the ATCA Images and PWN Flux Density Measurements \label{table:atcaimg}}
\tablehead{\colhead{Band} & \colhead{\texttt{robust}} &
\colhead{Beam size} & \colhead{rms noise} & \colhead{PWN flux}\\
& & \colhead{FWHM} & \colhead{(\mjb)} & \colhead{density (Jy)} }
\decimals
\startdata
20 cm & $-2$ & $23\arcsec\times20\arcsec$ & 0.06 & $0.17\pm0.01$ \\
13 cm & 0 & $15\arcsec\times13\arcsec$ & 0.11 & $0.14\pm0.01$ \\ 
6 cm & 0 & $16\arcsec\times14\arcsec$ & 0.04 & $0.08\pm0.01$ \\
3 cm & 0.5 & $13\arcsec\times12\arcsec$ & 0.04 & $0.06\pm0.01$ 
\enddata
\tablenotetext{}{The PWN flux density measurements correspond to regions 4 and 5 combined (shown in Figure \ref{fig-large-scale}).}
\end{deluxetable}


\section{RESULTS}

\subsection{Pulsar Motion}

Since the ATCA data have a large beam size ($>10\arcsec$), we used the {\sl CXO} data to search for changes in the pulsar position (and therefore, its proper motion).
The following procedure was performed to correct for systematic astrometric errors that may be present in the {\sl Chandra} World Coordinate System (WCS).

We ran {\tt wavdetect} (a Mexican-hat wavelet source detection algorithm; \citealt{Freeman2002}) on both observations. 
We excluded an $r=40''$ circle around the pulsar (to prevent nebular emission in the pulsar's vicinity from being misidentified as point sources), sources with $<$12 counts, and sources farther than 5$'$ from the optical axis (to filter out sources with poor localizations). 
We then ran {\tt wcs\_update} on both observations, using Gaia DR2 sources \citep{Bailer-Jones2018} as the reference source list, and set the {\tt radius} parameter to 0.8 (i.e., sources were considered a match if their optical and X-ray positions resided within $0\farcs5$ of each other).
Both observations had 8 source pairs (of which 3 source pairs were seen in both {\sl CXO} observations).
The best-fit frame shifts along (RA, Dec.)\ and their uncertainties were 
($118\pm 68$, $114\pm 66$) mas and ($-467\pm 92$, $67\pm60$) mas, for ObsIDs 3855 and 21357, respectively.
The {\sl CXO}-Gaia frame shift (transformation) uncertainty along RA or Dec.\ for a given CXO observation, $\sigma_{a}^{\rm tr}$,
is calculated from the equation
\begin{equation}
    (\sigma_{a}^{\rm tr})^{-2} = \sum_{i=1}^{N_a} (\sigma_{i,a})^{-2},
    \label{sigma-trans}
\end{equation}
where $a$ marks the {\sl CXO} observation, $N_a$ is the number of {\sl CXO}-Gaia pairs for this observation, and $\sigma_{i,a}$ is the uncertainty of $i$-th {\sl CXO} source coordinate along the chosen direction calculated by {\tt wavdetect}\footnote{See \url{https://cxc.harvard.edu/ciao/ahelp/wavdetect.html} for details.}.
The Gaia positional uncertainties are negligible compared to the {\sl CXO} ones.  

The transformations produced by {\tt wcs\_update} lowered the average offsets between the X-ray and optical positions of the sources, and these were used to update the aspect solutions of the {\sl Chandra} observations and register all the detected X-ray sources on the Gaia reference frame.

In each astrometrically-corrected {\sl CXO} observation we calculate the average position of all counts within $2\farcs5$ of the brightest pixel in the pulsar vicinity.
We find that the pulsar shifts by $ \Delta\alpha\,\cos\delta =-154\pm140$ mas and $ \Delta\delta =-440\pm116$ mas,
where $\alpha$ and $\delta$ are RA and Dec. 
The uncertainty of the pulsar shift in a given direction is obtained by summation in quadrature of the pulsar position uncertainties in two {\sl CXO} observations and two {\sl CXO}-Gaia transformation uncertainties (see Equation \ref{sigma-trans}).

Dividing the pulsar shifts over the time interval of 16.3 years between the {\sl CXO} observations, we obtain the pulsar proper motion
\begin{equation}
    \mu_\alpha = -9.4\pm 8.6\,{\rm mas\,yr}^{-1},\quad \mu_\delta = -26.9\pm 7.1\,{\rm mas\,yr}^{-1}.
\end{equation}
This corresponds to total proper motion $\mu=28.8\pm7.3$ mas yr$^{-1}$ oriented at a position angle $198^\circ\pm 17^\circ$ East of North. 
At distance $d=3.2$ kpc, this corresponds to a transverse pulsar velocity $v_\perp = 440 \pm 110$ km s$^{-1}$.

\subsection{PWN Morphology}
\begin{figure}
\includegraphics[width=1.0\hsize,angle=0]{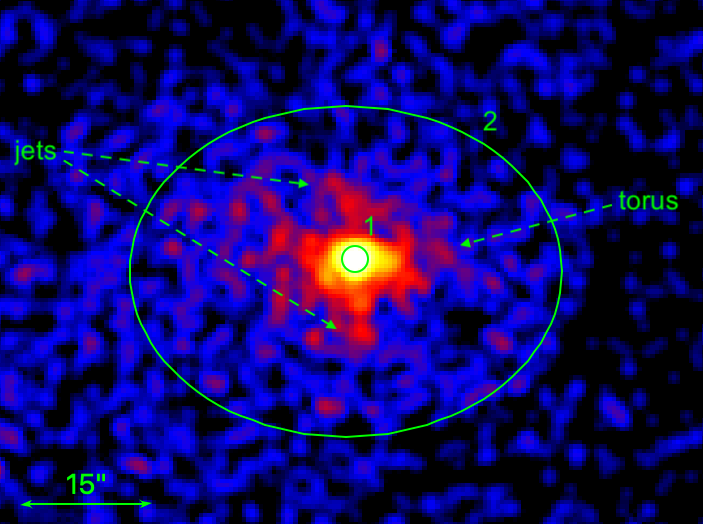}
\caption{Merged {\sl CXO} image of PSR J1016 (111.6 ks, 0.5--8~keV, smoothed with a 3-pixel ($r=1\farcs48$) Gaussian kernel), showing the small-scale structure.  The following regions are shown:  1 -- the pulsar (the $r=1\farcs5$ circle), and  2 -- the compact nebula (CN; the ellipse, excluding region 1).}
\label{fig-small-scale}
\end{figure}

In Figure \ref{fig-small-scale} we present the merged {\sl CXO} image showing the PWN's small-scale features in the vicinity of the pulsar.
The image reveals that the pulsar (region 1) embedded in a diffuse emission that could be interpreted as a torus and jets.
The putative torus/jets are also embedded within fainter diffuse emission.
Since, when fitted independently, the putative torus/jets and surrounding emission exhibited the same spectra, we defined both of these collectively as the compact nebula (CN; region 2).
No morphological changes in the PWN were seen across the two observations prior to producing the merged image.

\begin{figure*}
\includegraphics[width=1.0\hsize,angle=0]{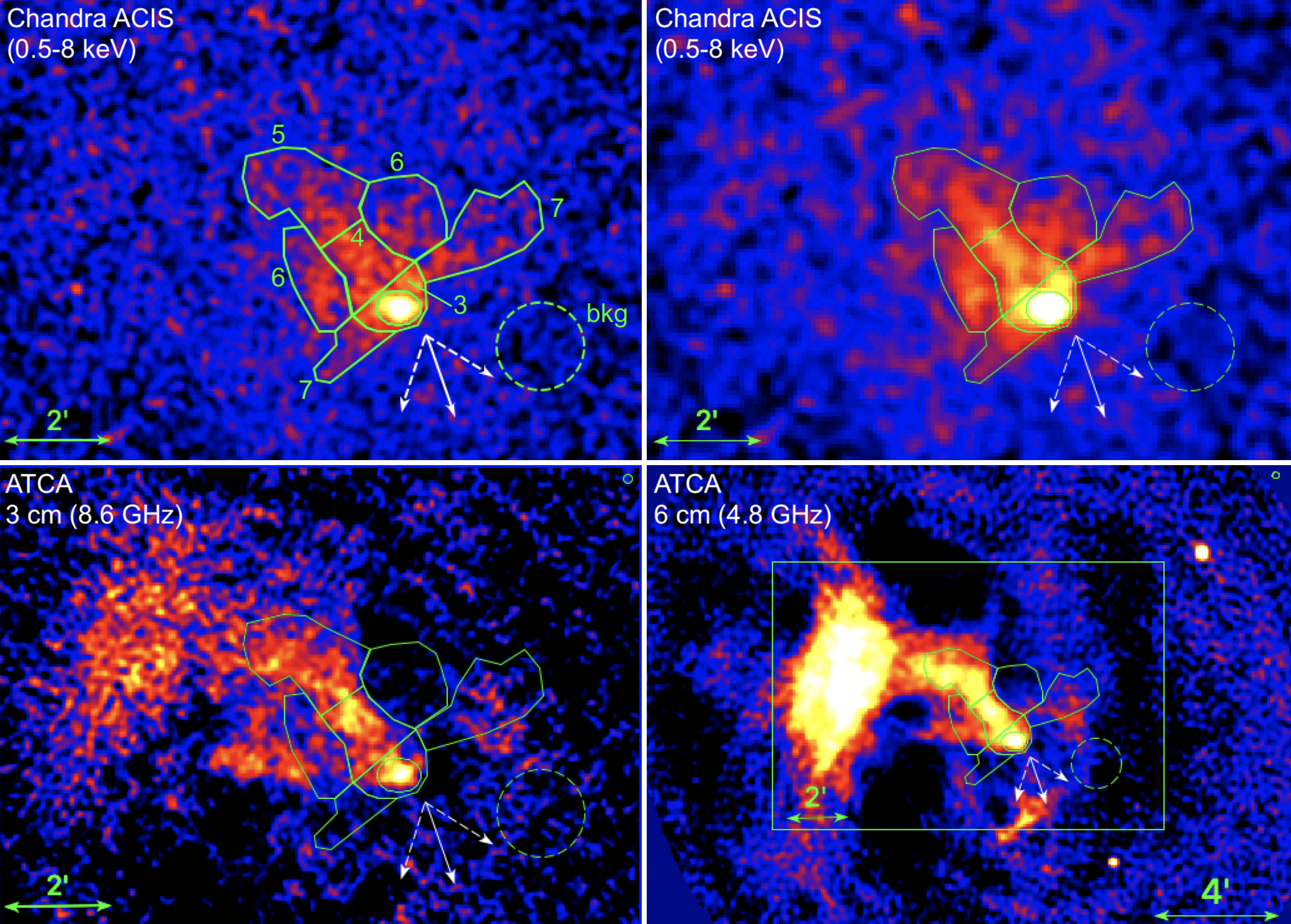}
\caption{{\sl Top:} Merged {\sl CXO} X-ray images of the J1016 PWN (with point sources removed).  
The images correspond to two different binnings/smoothings to roughly match the resolution of the ATCA radio images below them.  
The top-left panel is binned by a factor of 6 and smoothed with a 3-pixel ($r=8\farcs9$) Gaussian kernel, and the top-right panel is binned by a factor of 10 and smoothed with a 3-pixel ($r=14\farcs8$) Gaussian kernel.
The following regions used for X-ray spectroscopy are shown:  3 -- PWN head (excluding the region 2 ellipse), 4 -- tail (near), 5 -- tail (far), 6 -- lobes, 7 -- protrusions, and the region used for background subtraction (``bkg'').
The solid white arrow represents the direction of pulsar motion (see Section 3.1); the dashed white arrows represent the uncertainty (2$\sigma$) of the motion direction.
{\sl Bottom:} ATCA radio images of the J1016 PWN and surrounding field. 
The bottom-right panel is zoomed out; the green box represents the field of view shown in the other panels.
The ellipses in the top right corners of the radio images show the beam sizes: $8\farcs1\times7\farcs3$ for 3 cm, and $13\farcs7\times12\farcs5$ for 6 cm.}
\label{fig-large-scale}
\end{figure*}

\begin{figure*}[ht]
\epsscale{1.15}
\includegraphics[width=1.0\hsize,angle=0]{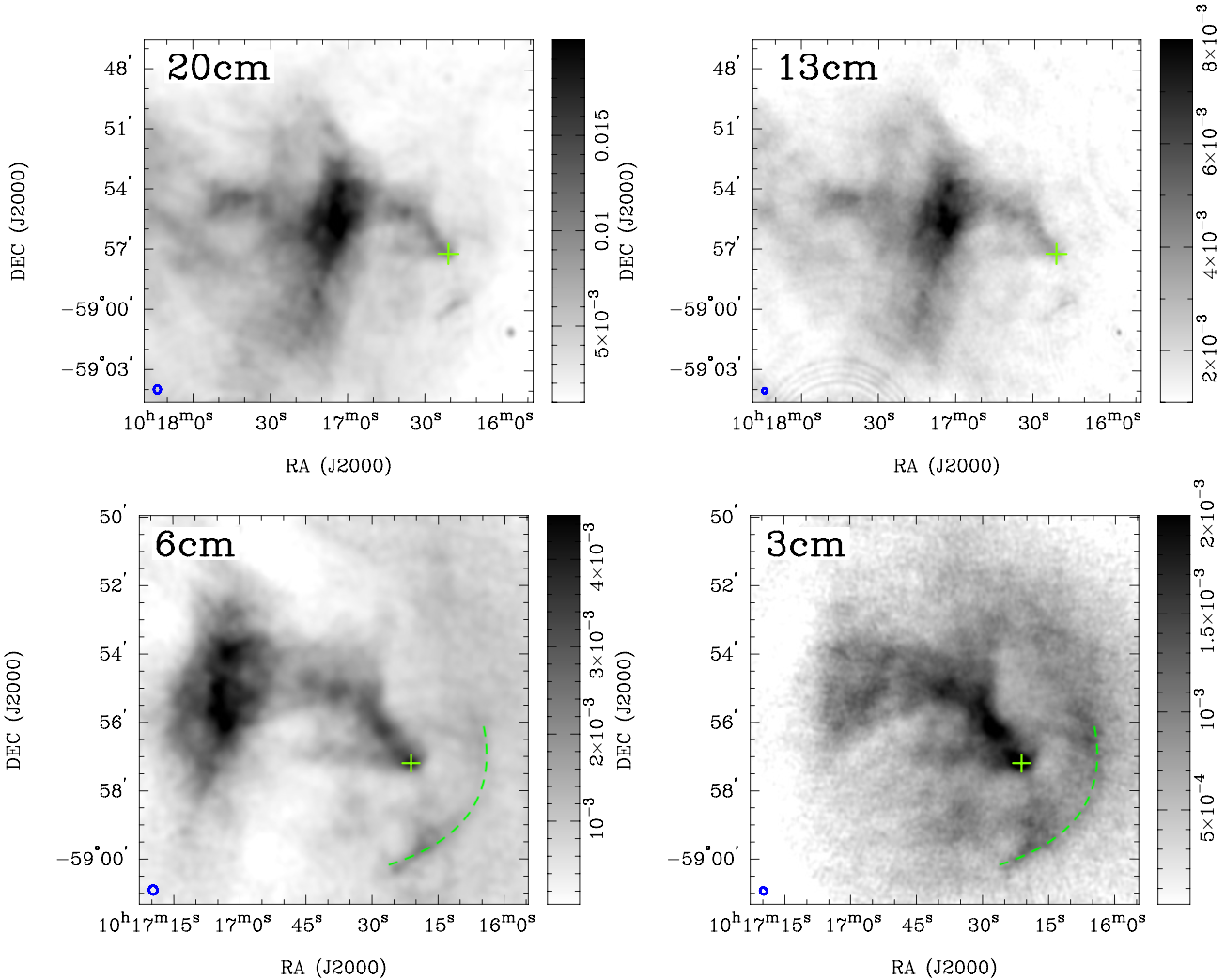}
\caption{Radio total intensity images of J1016 taken with ATCA.
The ``+'' sign marks the position of the pulsar and the beam size is shown in lower left.
The scale bars are in units of Jy\,beam$^{-1}$.
The dashed green lines mark the traces of shell-like emission seen.
\label{fig_atca}}
\end{figure*}

In Figure \ref{fig-large-scale} we present merged {\sl CXO} images and ATCA images (at 3 and 6 cm) of the J1016 PWN and its field.
In Figure \ref{fig_atca} we present all ATCA images (at 3, 6, 13, and 20 cm).
In radio, the PWN appears elongated to the Northeast, in a direction almost opposite that of the direction of pulsar motion: we interpret this emission as a pulsar tail (the extension labeled as the ``neck'' of ``the goose'' in Figure 1).
A narrow protrusion (somewhat fainter than the tail, and seen only in radio) extends about $3'$ eastward from the pulsar.
At roughly $2\farcm5$ NE of the pulsar, the radio tail (the neck of the goose) bends to the East.
Figure \ref{fig-multiwavelength} shows the wide-field SUMSS radio image of the complex J1016 field.
Roughly $4'$ East of the neck lies a large peculiarly-shaped structure (the body of ``the goose'').
The segment of radio emission after the bend in the neck extends through the goose body, up to 10$'$ to the West.
To the Southwest of the pulsar, in the 6 cm and 3 cm radio images (the bottom panels of Figure \ref{fig_atca}), traces of shell-like emission can be seen (which may be part of J1016's host SNR).

In X-rays, the PWN is brightest in the center of the tail along its axis (coincident with the radio emission), but also appears slightly wider than it does in radio (e.g., the ``lobes'': region 6 in Figure \ref{fig-large-scale}). 
These X-ray lobes appear to lack radio emission (although the radio protrusion passes through the eastern lobe).

\subsection{Radio Polarization}
\begin{figure}[hb]
\epsscale{1.15}
\plotone{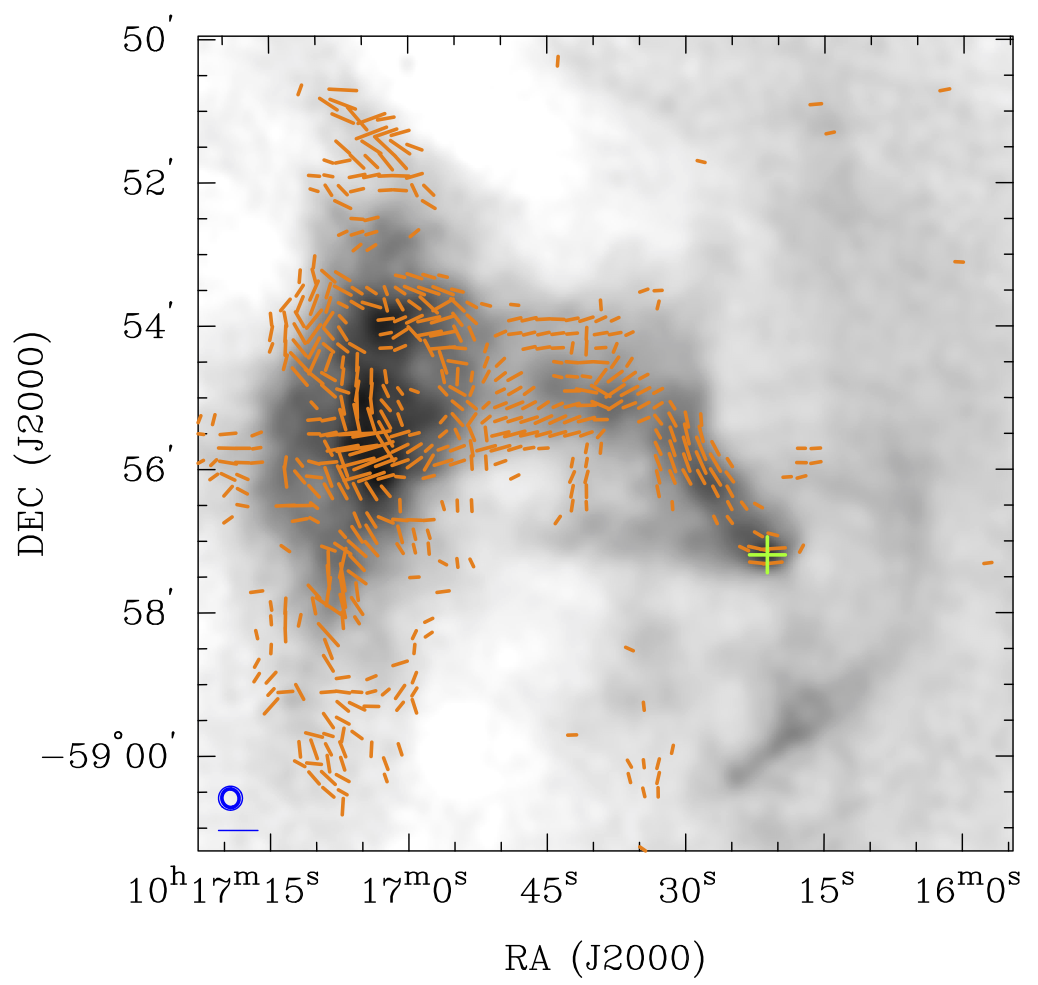}
\caption{6\,cm total intensity map of J1016 Figure~\ref{fig_atca},
overlaid with polarization $B$-vectors that indicate the intrinsic magnetic field orientation.
The vector lengths are proportional to the polarized intensity at 6\,cm. 
The vectors are clipped if the signal-to-noise ratio $<5$ in polarized intensity or $<10$ in total intensity,
or if the uncertainty in position angle $>20\arcdeg$.
The intensity and vector maps are smoothed to a resolution of 20\arcsec.
\label{fig_bfield}}
\end{figure}

To study the PWN's polarization, we focused on the 3 and 6\,cm maps, since they have better resolution and sensitivity than the lower frequency ones.
We first determine the foreground rotation measure (RM) using the polarization angles maps.
At the tip of the PWN, our RM map shows values that are fully consistent with that of the pulsar ($-540$\,rad\,m$^{-2}$). 
The RM increases gradually to $\sim-100$\,rad\,m$^{-2}$ along the pulsar tail.
We then used the RM map to correct for Faraday rotation of the polarization vectors; the intrinsic orientation of the PWN magnetic field is shown in Figure~\ref{fig_bfield}.
There is a good alignment between the magnetic field orientation and the axis of the pulsar tail.
However, at the neck, the orientation of the magnetic field appears to abruptly change by roughly 90$^\circ$.

\subsection{X-ray Spectra}

In order to find the best-fit value for the absorbing hydrogen column density $N_{\rm H}$, we first fit the spectrum from region 2, the CN (which excludes region 1, the pulsar; see Figure \ref{fig-small-scale}).
We selected this region because it is sufficiently bright and small enough that the effects of synchrotron cooling across its extent should be negligible.
Fitting with the absorbed power-law (PL) model, we found $N_{\rm H} = (0.94 \pm 0.18) \times10^{22}$ cm$^{-2}$ and $\Gamma_{\rm CN} = 1.77\pm0.16$, with $\chi^2_\nu = 0.96$ (for $\nu=35$ d.o.f.).
When fitting the CN and the pulsar simultaneously (allowing the photon indices to differ but linking $N_H$), we obtained a similar result, $N_{\rm H} = (0.91\pm0.20) \times 10^{22}$ cm$^{-2}$, $\Gamma_{\rm CN} = 1.75\pm0.16$, and $\Gamma_{\rm PSR} = 1.68\pm0.18$, with $\chi^2_{49} = 1.02$. 
The correlation between DM and $N_{\rm H}$ found 
by \citet{He2013}, $N_{\rm H} = 0.30_{-0.09}^{+0.13} \times 10^{20}\, {\rm DM}$ cm$^{-2}$, would suggest $N_{\rm H} \sim1.2\times10^{22}$ cm$^{-2}$ for J1016.
Our best-fit $N_{\rm H}$ is fairly close to this value. 
Thus, for all subsequent spectral analysis, we fix $N_{\rm H} = 0.91\times10^{22}$ cm$^{-2}$.

\begin{deluxetable*}{ccccccccc}
\tablecolumns{4}
\tablecaption{X-ray Spectral Fit Results for PWN Regions using an Absorbed PL Model}
\tablewidth{0pt}
\tablehead{\colhead{Region} & \colhead{Name} & \colhead{Area} & \colhead{Net Counts} & \colhead{$\Gamma$} & \colhead{$\mathcal{N}_{-5}$} & \colhead{$\chi_\nu^2$} & \colhead{$F_{-13}$} & \colhead{$L_{32}$} }
\startdata
1 & Pulsar & 7.1 & $453\pm21$ & $1.72\pm0.11$ & $2.23\pm0.27$ & 1.04 (26) & $0.78\pm0.04$ & $1.51\pm0.07$ \\ 
2 & Compact Nebula & 1,464 & $1320\pm41$ & $1.74\pm0.07$ & $5.58\pm0.69$ & 1.03 (55) & $1.91\pm0.08$ & $3.71\pm0.11$ \\
3 & PWN Head & 2,161 & $390\pm30$ & $2.48\pm0.18$ & $3.71\pm0.51$ & 1.09 (35) & $0.52\pm0.05$ & $1.55\pm0.12$ \\
4 & Tail (Near) & 4,863 & $606\pm46$ & $2.26\pm0.22$ & $4.36\pm0.83$ & 1.01 (40) & $0.79\pm0.10$ & $2.02\pm0.18$ \\
5 & Tail (Far) & 9,707 & $956\pm72$ & $2.08\pm0.19$ & $5.79\pm0.98$ & 1.23 (33) & $1.29\pm0.14$ & $2.99\pm0.24$ \\
6 & Lobes & 11,965 & $769\pm70$ & $2.25\pm0.19$ & $6.76\pm1.05$ & 1.41 (56) & $1.22\pm0.14$ & $3.15\pm0.24$ \\
7 & Protrusions & 10,926 & $599\pm67$ & $2.34\pm0.21$ & 8.51, 4.88 & 1.46 (62) & 1.39,  0.80 & 3.78, 2.17 \\
\hline
3-7 & PWN (minus CN) & 39,795 & $3369\pm226$ & $2.34\pm0.12$ & $33.0\pm2.8$ & 1.62 (109) & $5.43\pm0.24$ & $14.70\pm0.70$ \\
\enddata
\tablenotetext{}{Spectral fit results for the different regions of the PWN.  Listed are the region number, region name, area (in arcsec$^2$), net counts, photon index $\Gamma$, PL normalization $\mathcal{N}_{-5}$ in units of $10^{-5}$ photons s$^{-1}$ cm$^{-2}$ keV$^{-1}$ at 1 keV, reduced $\chi^2_{\nu}$ ($\nu$ d.o.f.), observed (absorbed) 0.5--8 keV fluxes $F_{-13}$ (in units of $10^{-13}$ erg cm$^{-2}$ s$^{-1}$), and luminosity $L_{32}$ (in units of $10^{32}$ erg s$^{-1}$).
In all fits we set $N_{\rm H}=0.91\times10^{22}$ cm$^{-2}$.
The ACIS chip gap crosses region 7 in ObsID 21357, so in the fits to this region we allow the normalizations between observations to vary; we list the resulting normalizations and fluxes for ObsIDs 3855 and 21357, respectively.
Note: The combined regions 3-7 data were better-fit with PL + thermal plasma models; see text for details.
}
\label{table-spectra}
\end{deluxetable*}

\begin{figure}
\includegraphics[width=1.0\hsize,angle=0]{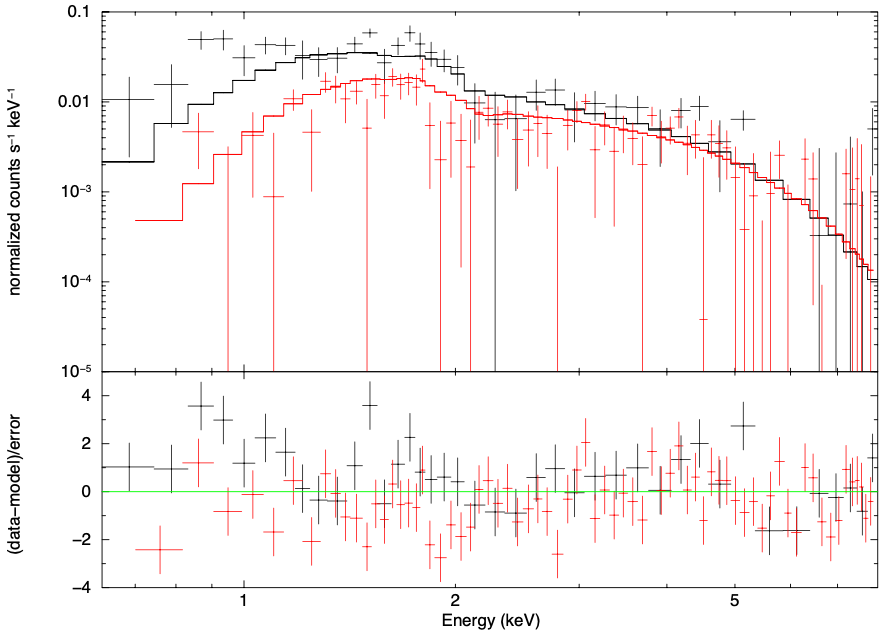}
\caption{Absorbed PL fit of regions 3-7 combined.  The black data points correspond to ObsID 3855, and the red data points correspond to ObsID 21357.  The fit details are provided in Table \ref{table-spectra}.  Note the systematic residuals seen below 1.5 keV in ObsID 3855.
These residuals are not seen in ObsID 21357 due to {\sl Chandra}'s loss of sensitivity to soft X-rays resulting from contamination accumulating on the ACIS optical blocking filters (see \citealt{Plucinsky2018}).}
\label{fig-regions-3-7-PL}
\end{figure}

In Table \ref{table-spectra} we list the spectral fit results for the pulsar and all regions of the PWN (shown in Figures \ref{fig-small-scale} and \ref{fig-large-scale}) fit individually.
For each region, we fit the spectra from the two {\sl CXO} observations simultaneously (rather than merge them, due evolution of the ACIS instrument response); no significant spectral changes are seen between the observations.
The pulsar and the CN exhibit virtually the same spectra, $\Gamma_{\rm PSR} = 1.72\pm0.11$ and $\Gamma_{\rm CN} = 1.74\pm 0.07$.
The rest of the PWN (regions 3-7) exhibit softer spectra, with photon indices in the range $\Gamma\approx 2.1-2.5$.
Since regions 3-7 exhibit similar spectra, we combined the regions, reextracted/refit the spectra, and obtained $\Gamma_{3-7} = 2.34\pm0.12$, though with a formally unacceptable (or rather large) reduced $\chi^2_{109} = 1.62$.
The fit is shown in Figure \ref{fig-regions-3-7-PL}.

\begin{figure}
\includegraphics[width=1.0\hsize,angle=0]{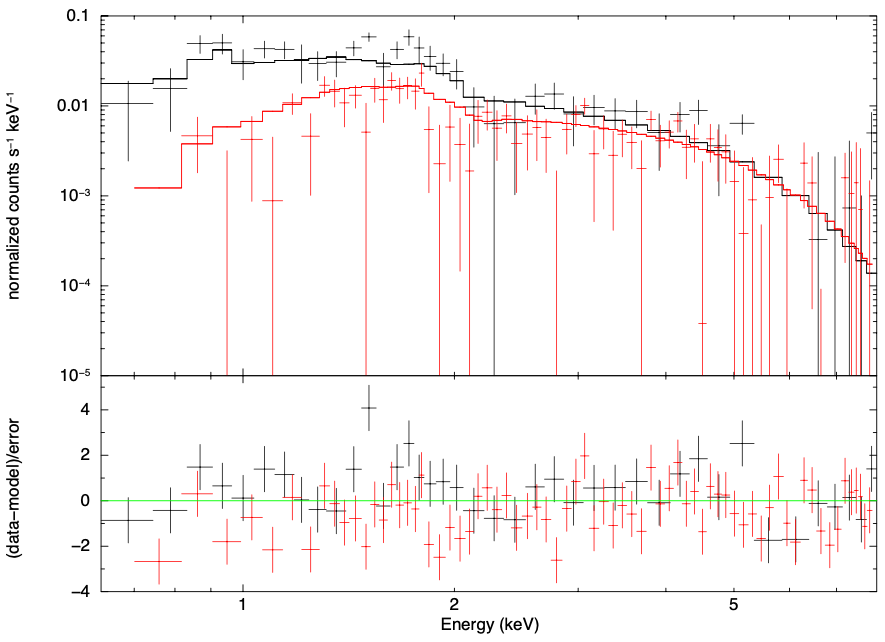}
\includegraphics[width=1.0\hsize,angle=0]{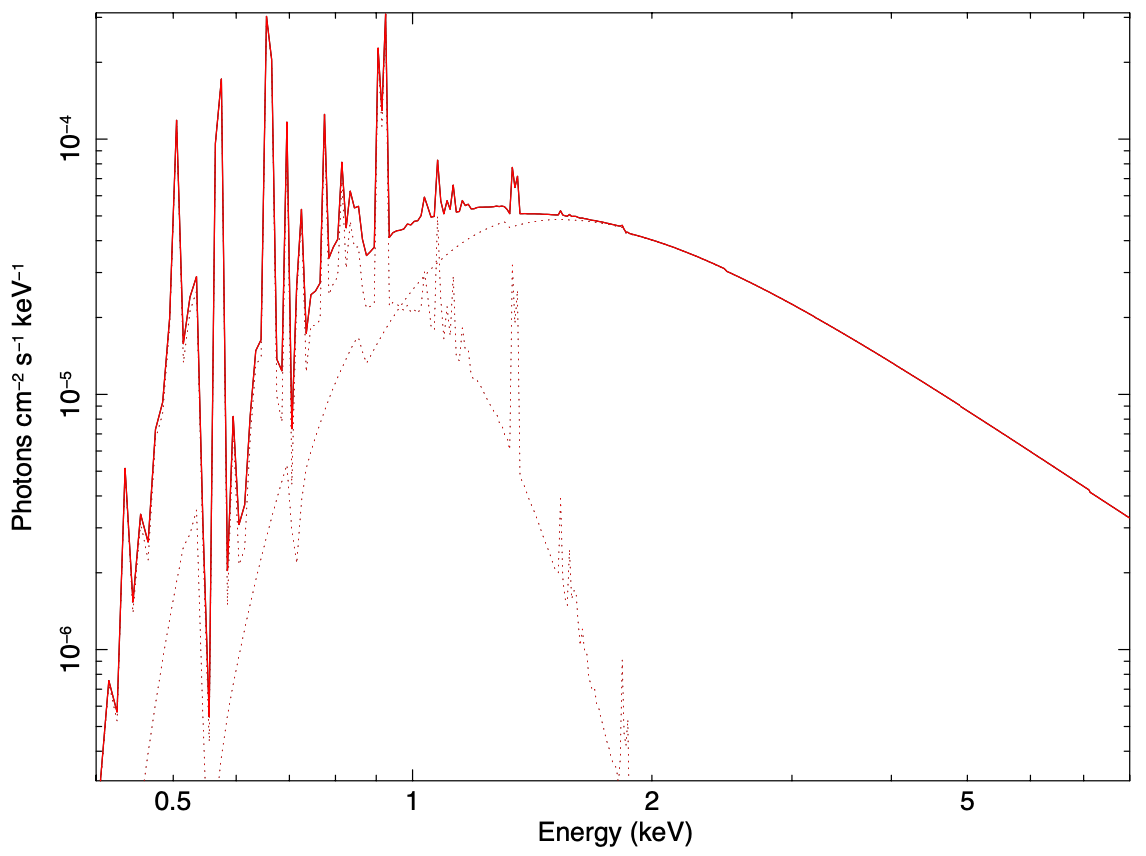}
\caption{Absorbed PL + {\tt apec} fit of regions 3-7 (top), and components in the absorbed photon spectrum (bottom).  In the fit (top), the black data points correspond to ObsID 3855, and the red data points correspond to ObsID 21357.  The fit details are provided in the text.}
\label{fig-regions-3-7-PL-plus-apec}
\end{figure}

The PL fit to regions 3-7 (Figure \ref{fig-regions-3-7-PL}) showed a significant data excess at low energies ($<2$ keV) in ObsID 3855 (which was taken before contamination accumulated on the ACIS detector and lowered its sensitivity to soft X-rays\footnote{See \url{https://cxc.cfa.harvard.edu/ciao/why/acisqecontamN0010.html}.}).
The excess could be due to soft emission, e.g., from a thermal plasma.   
Therefore, we tried fitting regions 3-7 with a PL plus emission from an optically thin, collisionally-ionized plasma in full thermal equilibrium (XSPEC's {\tt apec} model, assuming solar abundances) to account for the possibility that the pulsar is still residing in its progenitor SNR.
We obtained $kT = 0.14_{-0.04}^{+0.07}$ keV, {\tt apec} component normalization\footnote{\label{apec-norm} The {\tt apec} normalization is defined as $\mathcal{N}_{\rm apec} = 10^{-14}\, (4\pi D^2 )^{-1} \int n_{\rm e} n_{\rm H} dV$, where $D$ is the distance to the source (in cm), and $n_{\rm e}$ and $n_{\rm H}$ are the electron and Hydrogen number densities (cm$^{-3}$), respectively.} $\mathcal{N}_{\rm apec}=0.014_{-0.01}^{+0.07}$ cm$^{-5}$, $\Gamma=2.09\pm0.15$, and PL component normalization $\mathcal{N}_{\rm PL}=(2.57\pm0.37)\times10^{-4}$ photon s$^{-1}$ cm$^{-2}$ keV$^{-1}$ (at 1 keV), with $\chi^2_{107} = 1.44$.
This corresponds to an observed flux $F_{\rm 0.5-8\,keV} = (6.1_{-0.7}^{+0.1})\times10^{-13}$ erg cm$^{-2}$ s$^{-1}$, and an unabsorbed flux $F_{\rm 0.5-8\,keV}^{\rm unab} =(1.1\pm0.1)\times10^{-11}$ erg cm$^{-2}$ s$^{-1}$.
The observed fluxes for the PL and {\tt apec} components in the same energy range are $F_{\rm PL} = (5.5 \pm 0.3)\times 10^{-13}$ erg cm$^{-2}$ s$^{-1}$, and $F_{\rm apec} = (0.5 \pm 0.2)\times 10^{-13}$ erg cm$^{-2}$ s$^{-1}$.  
The unabsorbed fluxes for the PL and {\tt apec} components in the same energy range are $F_{\rm PL}^{\rm unab} = (1.1\pm0.1)\times10^{-12}$ erg cm$^{-2}$ s$^{-1}$, and $F_{\rm apec}^{\rm unab} = (9.5_{-1.7}^{+1.5})\times10^{-12}$ erg cm$^{-2}$ s$^{-1}$, respectively.
The fit is shown in Figure \ref{fig-regions-3-7-PL-plus-apec}.
We tried fitting the data with similar thermal equilibrium plasma models ({\tt mekal}, {\tt raymond}, and {\tt equil}) and obtained virtually the same fit parameters.

\begin{figure}
\includegraphics[width=1.0\hsize,angle=0]{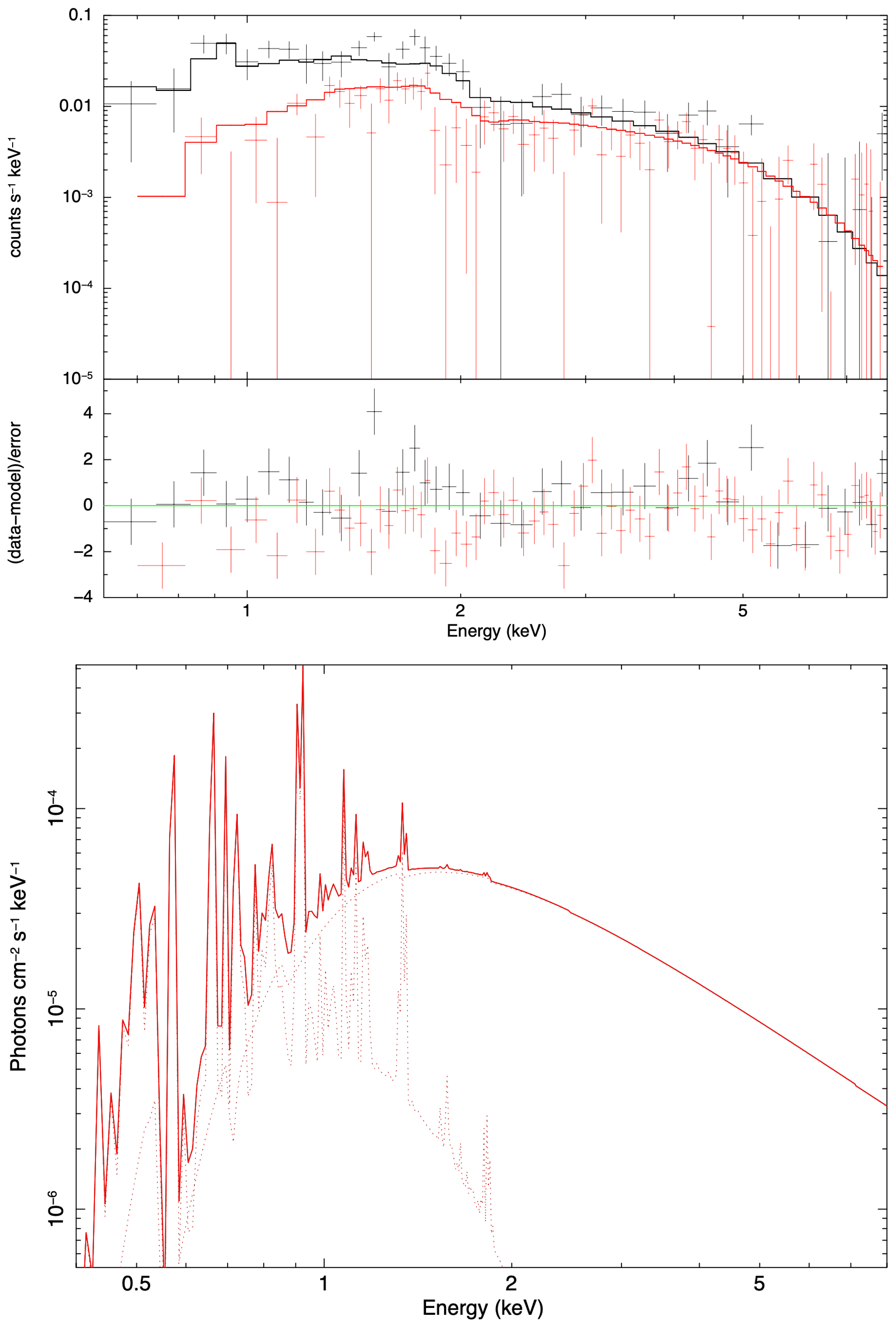}
\caption{Absorbed PL + {\tt nei} fit of regions 3-7 (top), and components in the absorbed photon spectrum (bottom).  In the fit (top), the black data points correspond to ObsID 3855, and the red data points correspond to ObsID 21357.  The fit details are provided in the text.}
\label{fig-regions-3-7-PL-plus-nei}
\end{figure}

Since J1016 is young, one can expect non-equilibrium ionization of the SNR plasma. 
To check this possibility, we fit the same spectra with a PL plus a model for emission from thermal plasma with non-equilibrium ionization (XSPEC's {\tt nei} model, assuming solar abundances).
The fit is shown in Figure \ref{fig-regions-3-7-PL-plus-nei}.
We obtained $kT = 0.27_{-0.12}^{+0.14}$ keV, ionization timescale $\tau = (1.2_{-1.0}^{+12})\times10^{10}$ s cm$^{-3}$, {\tt nei} normalization\footnote{{\tt Nei} normalization is defined by the same equation as the {\tt apec} normalization -- see footnote \ref{apec-norm}.} $\mathcal{N}_{\rm nei} = (2.0_{-1.4}^{+13})\times10^{-3}$, $\Gamma=2.1\pm0.2$, and PL normalization $\mathcal{N}_{\rm PL} = (2.6_{-0.6}^{+0.4}) \times10^{-4}$  cm$^{-2}$ keV$^{-1}$ at 1 keV, with $\chi^2_{106}=1.44$.
This corresponds to an observed flux $F_{\rm 0.5-8\,keV} = (6.1_{-1.6}^{+0.5})\times10^{-13}$ erg cm$^{-2}$ s$^{-1}$, and an unabsorbed flux $F_{\rm 0.5-8\, keV}^{\rm unab}= (9.2\pm0.4)\times10^{-12}$ erg cm $^{-2}$ s$^{-1}$.
For the PL and {\tt nei} components, the unabsorbed fluxes are $F_{\rm PL}^{\rm unab} = (1.1\pm0.1)\times10^{-12}$ erg cm$^{-2}$ s$^{-1}$ and $F_{\rm nei}^{\rm unab} = (8.2_{-1.5}^{+1.4})\times10^{-12}$ erg cm$^{-2}$ s$^{-1}$, for the same energy range.
The observed fluxes for the PL and {\tt nei} components are $F_{\rm PL} = (5.6\pm0.8)\times10^{-13}$ and $F_{\rm nei} = (3.8\pm0.7)\times10^{-14}$ erg cm$^{-2}$ s$^{-1}$.
Thus, the data can be described equally well by adding to the PL component a thermal component emitted from either a plasma in full collisional equilibrium or a plasma with non-equilibrium ionization.
The data are not of high enough quality to allow fitting of elemental abundances.

\begin{figure}
\includegraphics[width=1.0\hsize,angle=0]{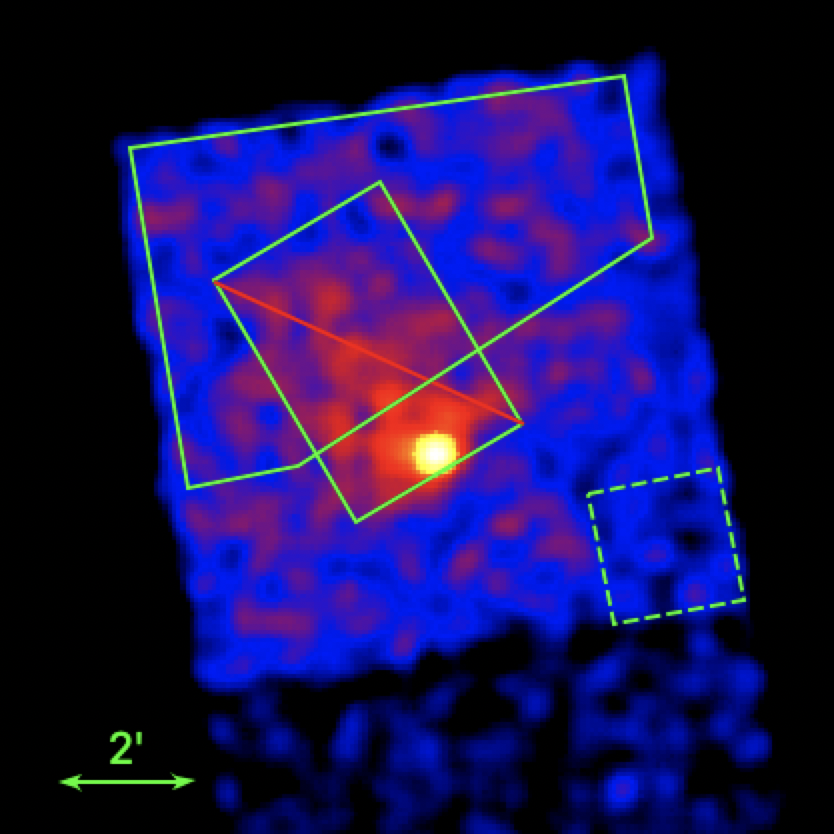}
\caption{{\sl CXO} ObsID 3855 (binned by a factor of 8, and smoothed with a $r=20''$ (5-pixel) Gaussian kernel to highlight the extended emission), showing the ``PWN surroundings'' region (solid polygon, minus the exclusion rectangle) and corresponding background region (dashed square).}
\label{fig-pwn-surroundings}
\end{figure}

To investigate the thermal emission seen in the PWN, we extracted the spectrum from the area surrounding the PWN (the region used is shown in Figure \ref{fig-pwn-surroundings}).
For this analysis we use only ObsID 3855, since ObsID 21357 was taken when ACIS's sensitivity to soft X-rays has been heavily degraded by the contamination accumulating on ACIS's optical blocking filter, and thus, is not very useful in probing soft thermal emission.
We find that the emission is best-fit by an absorbed PL + {\tt apec} model, with $\Gamma=1.82\pm0.37$, $\mathcal{N}_{\rm PL} = (1.85\pm0.62)\times10^{-4}$ photon s$^{-1}$ cm$^{-1}$ keV$^{-1}$ (at 1 keV), $kT = 0.19\pm0.06$ keV, and $\mathcal{N}_{\rm apec} = (8.1\pm1.0)\times10^{-3}$, with $\chi^2_{62} = 0.92$.
For comparison, fitting the same region with a PL-only model yielded $\Gamma=2.92\pm0.33$ with $\chi^2_{64} = 1.25$, and an {\tt apec}-only model yielded $kT = 0.74 \pm 0.09$ keV with $\chi^2_{64} = 1.30$.
Thus, the area surrounding the visible extent of the PWN appears to also be a mixture of thermal plasma and nonthermal electrons.

\section{DISCUSSION}

The X-ray spectroscopy of the tail and its surroundings revealed the presence of thermally-emitting plasma in the J1016 field.
This, as well as the nearby structures seen in the radio images (e.g., the traces of shell-like emission seen southwest of the pulsar in the bottom panels of Figure \ref{fig_atca}), suggests that the pulsar is still inside its progenitor SNR.
The boundaries of the SNR are not fully detected, which is somewhat surprising considering the pulsar's young spin-down age of 21 kyr (though the true age is likely even smaller; see \citealt{Igoshev2020}).

With the best-fit ${\tt apec}$ normalization $\mathcal{N}_{\rm apec}=0.014_{-0.01}^{+0.07}$ cm$^{-5}$ from the combined regions 3-7 fit, we can crudely estimate the Hydrogen number density within the PWN $n_H \sim 6-13$ cm$^{-3}$.
This estimate assumes a large ionization fraction ($n_H \sim n_e$) and approximates the pulsar tail as a cylinder of $l=3'$ and $r=0.5'$ at $d=3.2$ kpc ($V=5.6\times10^{55}$ cm$^3$).
Thus, the emission measure is crudely $\sim 5.6\times10^{57}$ $(d/3.2\ {\rm kpc})^2$ cm$^{-3}$.
With the best-fit temperature, $T=1.6_{-0.5}^{+0.8}$ MK, the above result implies pressure $P\sim(2-8)\times10^{-9}$ dyne cm$^{-2}$.

The nearby SNR G284.3--1.8 (the center of which is located $\sim$20$'$ to the East) may not associated with the high-mass X-ray/gamma-ray binary 1FGL J1018.6--5856 \citep{Marcote2018}. 
The similar observed $N_{\rm H}$ of SNR G284.3--1.8 and PSR J1016--5857 seems to suggest an association ($N_{\rm H,G284} = (0.91\pm0.09)\times10^{22}$ cm$^{-2}$ and $N_{\rm H,J1016} = (0.91\pm0.20)\times10^{22}$ cm$^{-2}$; see \citealt{Williams2015}).
If PSR J1016 and SNR G284.3 were associated, it would require the pulsar's transverse velocity $v_\perp \sim900$ km s$^{-1}$, and we would see a long tail extending eastward (which we do not see).
Also, J1016's velocity vector does not seem to point at (or near) the SNR center.
Thus, we consider the association between J1016 and G284.3 unlikely.

The tail-like morphology of the PWN seen in the radio and X-ray images suggests the confinement of the pulsar wind by the ram pressure due to the pulsar's motion through the ambient medium.
The direction of proper motion is generally in agreement with the shape of the PWN and the direction of the tail (see Figure \ref{fig-large-scale}).
At J1016's DM distance $d=3.2$ kpc, the pulsar motion measured from the {\sl CXO} data corresponds to a velocity $v_\perp = 440\pm110$ km s$^{-1}$, which is typical for pulsars with measured proper motion \citep{Verbunt2017}.

The presence of small-scale structures in the pulsar vicinity (i.e., the tentative torus/jets seen in Figure 2) would suggest transonic pulsar motion with a modest Mach number, $\mathcal{M} \equiv v / c_s \sim 1$, where $v$ is the pulsar velocity with respect to the ambient medium, and $c_s$ is the speed of sound in this medium; at higher Mach numbers such structures would be crushed by the ram pressure and be indiscernible (cf.\ images of high Mach number pulsars in \citealt{Kargaltsev2017b}).
However, the elongated radio PWN morphology argues for supersonic motion.
One can estimate the speed of sound in the pulsar's vicinity as $c_s \sim 150\  (\mu/0.6)^{-1/2}\ T_6^{1/2} $ km s$^{-1}$, where $\mu$ is the molecular weight and $T_6$ is the temperature in units of MK.
Using the temperatures obtained from the above fits, {\tt apec} and {\tt nei} respectively, we find $c_s\ \sim 190$ km s$^{-1}$ and $c_s\sim 270$ km s$^{-1}$.
These suggest that the pulsar is mildly supersonic and is still moving within the SNR interior. 
However, if the pulsar is indeed 21 kry old, it should have moved by about $10'$ during its lifetime. 
For a SNR radius of $10'$, the usual nominal Sedov age estimate gives a SN age  $t_{\rm SN}\approx16 (d/5~{\rm kpc})^{5/2}(n/E_{51})^{1/2}$ kyrs where $n$ cm$^{-3}$ is the local ISM density and $10^{51}E_{51}$ erg is the SN explosion energy.
This suggests that the true pulsar age may be smaller than the spin-down age unless the local ISM density is high or the SN explosion had a low yield.

The X-ray and radio images link the pulsar/PWN to the remarkable radio structure, ``the Goose'' (see the bottom left panel of Figure \ref{fig-multiwavelength}).
The body of the Goose seen in radio could be interpreted as parts of the PWN and/or SNR displaced by a reverse shock passage in the SNR (e.g., analogous to the bright radio filament in the Vela-X complex; see \citealt{Slane2018} and references therein). 
The broken shape of the radio/X-ray tail (i.e., the sharp bend separating the goose's ``neck'' and ``body'') can be explained by the reverse shock passing from the southwest to the northeast (more specifically, the reverse shock would be moving inward spherically toward the SNR center).

The spectra extracted from the tail's surroundings reveal that this region is a mixture of thermal plasma and relativistic electrons from the PWN.
This indicates that the pulsar wind particles are not entirely confined to the tail.
These particles could have either been displaced by the reverse shock interaction, diffused out of the tail (depending on its evolutionary stage), or leaked out via reconnection between the tail and ambient magnetic fields (see, e.g., \citealt{Bandiera2008,Barkov2019,Olmi2019a,Olmi2019b}).
This result further indicates that the PWN resides within a SNR.

\begin{figure}
\includegraphics[width=1.0\hsize,angle=0]{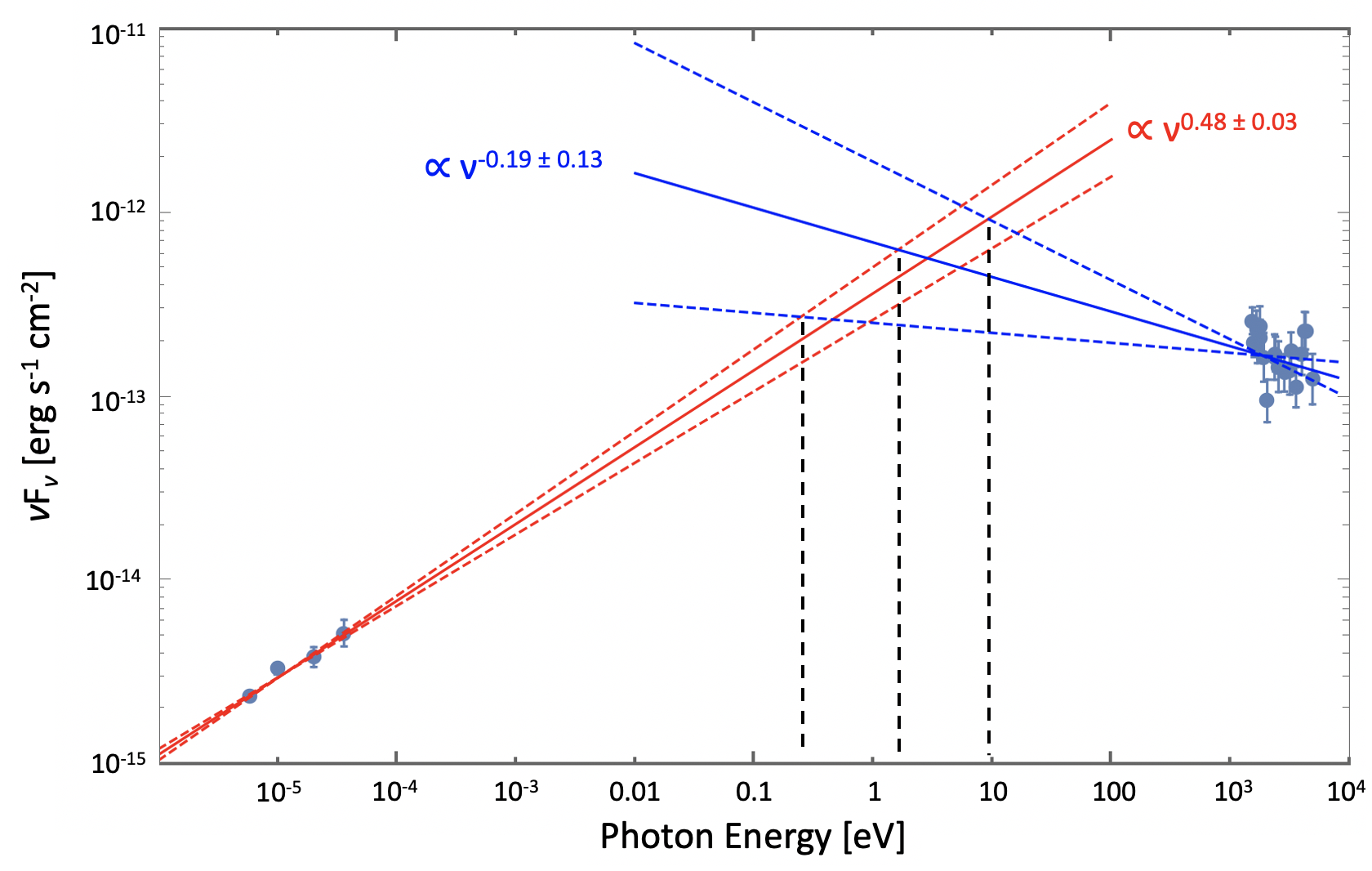}
\caption{Multiwavelength spectrum of the J1016 pulsar tail (regions 4 + 5).  The solid red and blue lines mark the PL slopes obtained from the radio and X-ray data, respectively, with their 1$\sigma$ uncertainties shown by the dashed lines.  The vertical black dashed lines mark the range of possible locations of the spectral break (assuming one spectral break between the radio and X-ray spectra). }
\label{fig-multiwavelength-spectrum}
\end{figure}

In Figure \ref{fig-multiwavelength-spectrum} we plot the multiwavelength spectrum of the pulsar tail (regions 4 + 5).
The radio spectrum is best-fit with a PL having photon index $\Gamma_R=1.52\pm0.03$ (or $\alpha_R=0.48\pm0.03$, where $\nu F_\nu \propto \nu^\alpha$), and the X-ray spectrum is best fit by a PL with $\Gamma_X=2.19\pm0.13$ (or $\alpha_X=-0.19\pm0.13$).
The difference in spectral slopes $\Delta\alpha = \alpha_R -\alpha_X = 0.67\pm 0.16$.
This result is generally in agreement with what one would expect from synchrotron cooling considerations, which predicts $\Delta\alpha =0.5$.
Deviations from $\Delta\alpha = 0.5$ are not uncommon in PWNe \citep{Chevalier2005,Reynolds2017}, and may indicate the presence of additional mechanisms, such as entrainment (mass loading of the ISM; \citealt{Morlino2015}), turbulent magnetic field amplification, diffusion, and/or particle reacceleration via magnetic reconnection \citep{Xu2019}.

The radio and X-ray measurements suggest that the spectrum should exhibit at least one spectral break between $3.6\times10^{-5}$ eV (8.7 GHz) and 0.5 keV.
Assuming only one spectral break, the best-fit radio/X-ray slopes suggest that it should occur around 3 eV. 
The radio spectrum does not rise as steeply as that of the Mouse PWN (one of the few pulsar tails bright in both radio and X-rays; \citealt{Klingler2018}), where, as a result, the break would occur at a significantly lower frequency. 
For the Mouse PWN, a double break may be more likely (Figure 11 in \citealt{Klingler2018}), which does not seem to be required by the current data for the J1016 tail, but which also can not be excluded. 
It is unclear what causes radio spectra to have different slopes, as synchrotron self-absorption effects are unlikely at $\nu >1$ GHz for either of the two PWNe. 
It is possible that different spectral slopes are caused by radiating electron populations with differing SEDs, but that would prompt the question of what causes the electron populations' spectra to differ.

The multiwavelength spectrum of the tail shown in Figure \ref{fig-multiwavelength-spectrum} indicates a spectral break at a frequency $\nu_c$ between the radio and X-ray frequencies. 
The observed change in the $\nu F_\nu$ spectral slope is consistent (within the measurement uncertainties for the slopes of radio and X-ray spectra) with a cooling break causing a spectral index change $\Delta\alpha =0.5$. 
The measurement uncertainties also imply that the break (assuming it is a single break) occurs  at $h\nu_c$ between 0.3 eV and 10 eV (see the dashed lines in Figure \ref{fig-multiwavelength-spectrum}), but likely closer to $h\nu_c = 0.3$ eV to give $\Delta\alpha =0.5$ -- a  canonical value for an optically-thin synchrotron spectrum in the slow cooling regime (see, e.g., \citealt{Klingler2018}, for a more detailed discussion). 
In this scenario, the slope $p$ of the uncooled part of the electron PL SED can be obtained from the observed radio spectrum as $p \equiv 3-2\alpha_R = 2.04\pm 0.06$.
With this slope, Equation (B16) from \citet{Klingler2018} with $\nu_c = 7.25\times10^{13}$ Hz ($=0.3$ eV), $\nu_m=843$ MHz, $\nu_M = 1.9\times10^{18}$ Hz  ($8$ keV\footnote{The actual values of $\nu_m$ and $\nu_M$ are not known, but the estimate is insensitive to $\nu_M$ for $p=2.04$}), $\nu_1 = 1.2\times10^{17}$ Hz ($=0.5$ keV), and $\nu_2 = \nu_M = 1.9\times10^{18}$ Hz ($=8$ keV) yields $B\sim (50-60) \sigma^{2/7}$~$\mu$G, with the range reflecting a weak dependence on the unknown $\nu_m$ which is assumed to be in the range of $0.01-1000$ MHz for the above estimate.
Here, $\sigma$ is the magnetization of the wind, frequencies $\nu_m$ and $\nu_M$ represent the minimum and maximum (respectively) synchrotron frequencies of the injected electron SED, $\nu_1$ and $\nu_2$ represent the boundary synchrotron frequencies of the observed band, and $\nu_c$ is the cooling frequency (i.e., the spectral break frequency).

\section{Conclusions}

The morphology of the X-ray PWN revealed by the new {\sl CXO} observations matches well the radio PWN morphology behind the moving pulsar (i.e., the pulsar tail).
At larger distances, the pulsar tail fades in X-rays, but the radio emission becomes brighter.
About 3$'$ NE of the pulsar, the tail abruptly bends and appears to connect with a larger radio structure (possibly the relic PWN; ``the Goose''). 
We attribute this to an interaction with the reverse shock inside the PWN's host SNR.
We measure the pulsar's proper motion, $\mu=28.8\pm7.3$ mas yr$^{-1}$ (at a position angle $198^\circ \pm 17^\circ$ East of North), which corresponds to projected velocity $v_{\perp} = 440\pm110$ km s$^{-1}$ (at $d=3.2$ kpc).
The spectroscopy of the PWN and its vicinity indicates the presence of a thermal plasma, providing further evidence that the PWN still resides within its host SNR.
We obtain the multiwavelength spectrum of the pulsar tail and estimate a magnetic field $B\sim(50-60)\sigma^{2/7}$ $\mu$G.
The relic PWN is expected to be a TeV source, which may be resolved with the Cherenkov Telescope Array (CTA) from the adjacent brighter H.E.S.S.\ source.

\facility{{\sl CXO}, ATCA} 
\software{CIAO v4.12, XSPEC v12.11.1, MIRIAD}

\acknowledgements
Support for this work was provided by the National Aeronautics and Space Administration through {\sl Chandra} award no.\ G09-2066 issued by the {\sl Chandra} X-ray Center, which is operated by the Smithsonian Astrophysical Observatory for and on behalf of the National Aeronautics and Space Administration under contract NAS8-03060.
JH acknowledges support from an appointment to the NASA Postdoctoral Program at the Goddard Space Flight Center, administered by the USRA through a contract with NASA.

The authors wish to thank the anonymous referee for their careful reading and helpful comments which have enhanced the clarity of this paper.

\end{document}